Editors' Suggestion

# Dislocation-driven nucleation type switching across repeated ultrafast magnetostructural phase transition

Jan Hajduček,[1,*] Antoine Andrieux,[2] Jon Ander Arregi,[1] Martin Tichý,[3] Paolo Cattaneo,[2] Beatrice Ferrari,[4] Fabrizio Carbone,[2] Vojtěch Uhlíř,[1,3] and Thomas LaGrange[2,†]

[1]CEITEC BUT, Brno University of Technology, Purkyňova 123, 612 00 Brno, Czech Republic

[2]Institute of Physics (IPHYS), Laboratory for Ultrafast Microscopy and Electron Scattering (LUMES), École Polytechnique Fédérale de Lausanne (EPFL), Lausanne 1015 CH, Switzerland

[3]Institute of Physical Engineering, Brno University of Technology, Technická 2, 616 69 Brno, Czech Republic

[4]Università degli Studi di Milano-Bicocca, Piazza dell'Ateneo Nuovo, 1-20126 Milano, Italy

Controlling magnetic order on ultrafast timescales, driven by spintronic and recording applications, is one of the main directions of current research in magnetism. Despite major advances in understanding the temporal evolution of magnetic order upon its emergence or quenching, experimental demonstration of the local link between microstructure and dynamic nucleation is missing. Here, taking advantage of the high structural and magnetic resolution of *in situ* transmission electron microscopy, we observe that cumulative laser irradiation significantly alters the nucleation pathway of the first-order antiferromagnetic to ferromagnetic phase transition of FeRh thin films, causing the transition to switch from homogeneous to heterogeneous nucleation. This leads to a decrease of 20 K in transition temperature and the emergence of submicron magnetic vortices as preferential nucleation motifs. These vortices are pinned in the film by underlying dislocation networks. We observe that the dislocation networks are formed and rearranged upon repeated crossing of the phase transition using both femtosecond and picosecond laser pulses. Our results establish a direct link between defect formation and the microscopic morphology of the nucleated ferromagnetic phase, with broad implications for ultrafast stroboscopic experiments and defect-mediated phase transitions in functional materials.

## I. INTRODUCTION

Disorder and dislocations are known to modulate phase transitions in functional materials, especially those undergoing first-order phase transitions (FOPTs), where the inherent phase coexistence sensitizes nucleation and growth dynamics to local heterogeneities [1–6]. In complex, strongly correlated systems, such as phase-change materials with coupled structural, electronic, or magnetic order, defects not only perturb the local lattice symmetry, but also directly influence emergent functionalities [7–10]. However, despite growing interest in defect engineering for tuning phase transition behavior, the microscopic role of dislocations in controlling the earliest stages of phase nucleation remains elusive.

This issue is even more relevant in nonequilibrium and laser-driven phase transitions, where repeated stimuli can dynamically create, accumulate, or stabilize defects [11–14]. The evolving defect landscape can, in turn, modify the phase transition pathway, introducing potential irreversibilities that would remain hidden in time-resolved stroboscopic studies due to limited spatial resolution. Understanding how repeated switching cycles reshape the local microstructure and how this alters phase nucleation kinetics is thus crucial not only for robust device design, but also for accurate interpretation of ultrafast measurements in correlated systems.

This work focuses on FeRh as a prototypical material to investigate these effects. The equiatomic, B2-ordered alloy exhibits a sharp FOPT from antiferromagnetic (AF) to ferromagnetic (FM) order just above 360 K, accompanied by a 1% lattice volume expansion, decrease in resistivity, and increase in entropy [15–17] [see Fig. 1(a)]. The transition can be triggered thermally, optically, or by strain, and is well studied on multiple timescales and using different probes [18]. Several studies have shown that the structural disorder induced in FeRh films by ion irradiation [19–21] or laser irradiation [22] can tune the magnetic transition temperature or even locally "write" FM regions [23].

Furthermore, numerous works have investigated the laser-induced AF-to-FM phase transition in FeRh to gain a more fundamental understanding of the coupled ultrafast electronic, structural, and magnetic transitions [24–31]. While these studies emphasize either the collective influence of disorder on magnetic ordering or the temporal dynamics of the laser-driven transition, the local mechanisms by which dislocations and their strain fields mediate phase nucleation remain largely unexplored. As a result, important aspects of how defects couple nanometer-scale domain formation with topology and influence transition dynamics under cyclic laser excitation across the AF-FM transformation may be overlooked.

*Contact author: jan.hajducek@ceitec.vutbr.cz
†Contact author: thomas.lagrange@epfl.ch

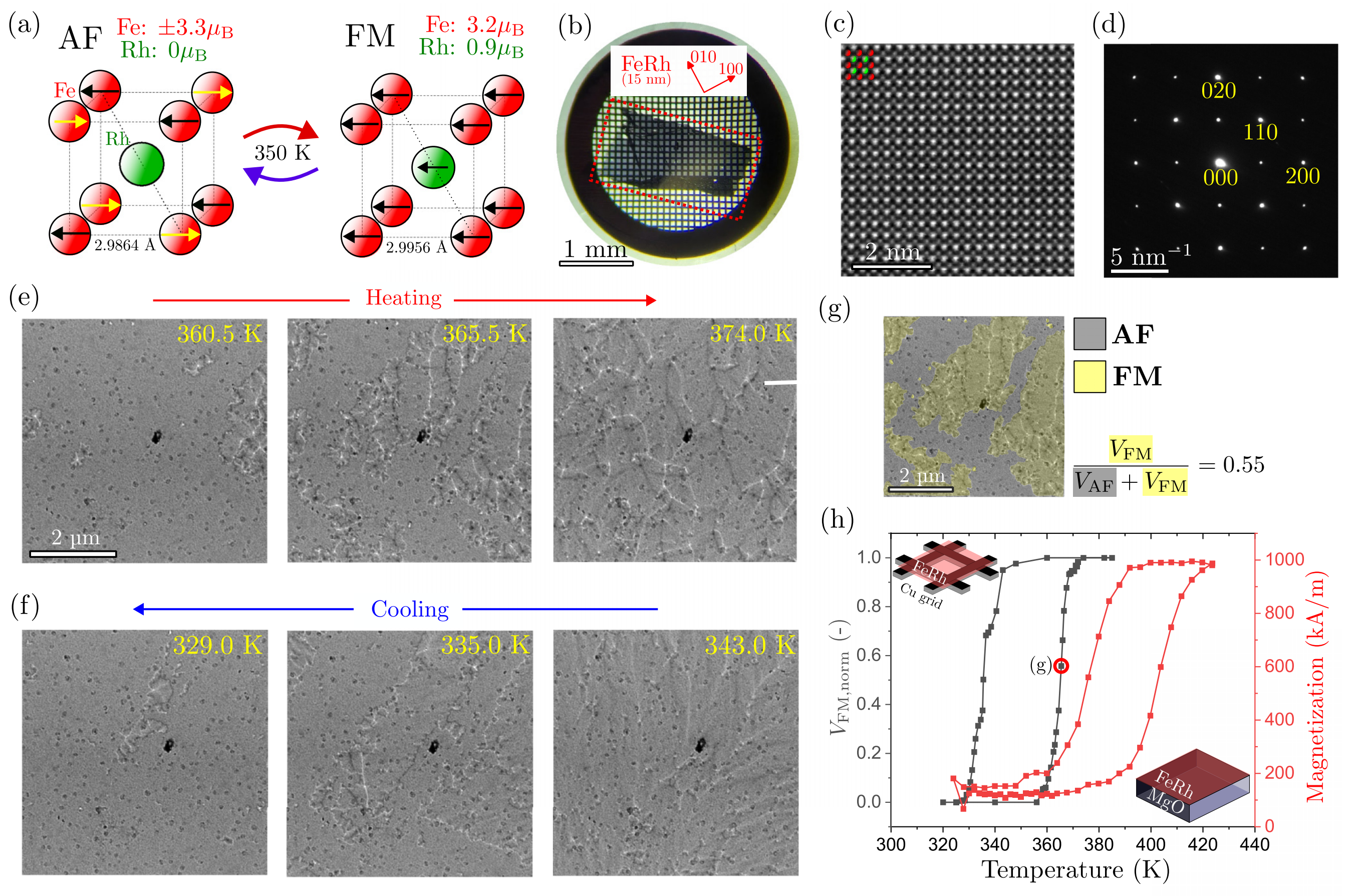

FIG. 1. Magnetostructural phase transition in FeRh. (a) Scheme of the magnetic phases in FeRh. (b) Photo of the freestanding FeRh film on a Cu TEM grid. (c), (d) HR-STEM high-angle annular dark-field image and SAED pattern of the suspended FeRh film along the [001] zone axis, respectively. (e), (f) Temperature-dependent series of LTEM images showing the gradual increase and decrease of the FM phase fraction during heating and cooling, respectively. (g) LTEM image-based quantification of the FM phase fraction at 365.5 K during heating. (h) Temperature-dependent FM phase fraction estimated from LTEM imaging (black curve) and magnetization measurements (red curve) for the suspended and supported FeRh films, respectively, showing the thermal hysteresis associated with the FOPT. LTEM data are locally obtained on the freestanding FeRh, while the magnetometry data are averaged from the full film on the substrate.

Here, we implement an *in situ* methodology that combines high-spatial-resolution imaging with controlled, repeatable laser excitation protocols. By using cumulative femtosecond laser exposure inside a transmission electron microscope (TEM), we directly observe how the magnetic microstructure of the product FM phase evolves in freestanding FeRh thin films under repeated, nonequilibrium stimuli using an *in situ* Lorentz TEM (LTEM). The freestanding geometry facilitates strain relaxation and provides an ideal platform for examining the interplay between lattice defects, strain, and magnetic phase formation with nanoscale precision. In particular, with specific laser treatment, we can program the FeRh films to exhibit a heterogeneous nucleation behavior that produces magnetic vortex textures as the dominant magnetic motif, which is of great interest both fundamentally and for technological applications. Our spatially resolved analysis provides insights into the microscopic mechanisms governing phase transitions in magnetic materials, highlighting the importance of defect engineering in the control of nanoscale phases.

## II. QUASISTATIC THERMALLY DRIVEN PHASE TRANSITION

To study the emergence of the FOPT in FeRh, we first conducted magnetic imaging during *in situ* Joule heating via the sample stage within the TEM. 15-nm-thick FeRh films were epitaxially grown on MgO(001) substrates via magnetron sputtering. Subsequently, the films were detached from the substrate by etching MgO in a mild acid solution [32,33] and transferred to a Cu TEM grid [34] (see Appendix A). This approach allows obtaining millimeter-sized freestanding FeRh flakes, such as the one shown in Fig. 1(b). The flakes preserve the high-quality crystalline order of the epitaxial, as-grown film, as apparent from the high-resolution scanning TEM (HR-STEM) and the selected area diffraction (SAED) images in Figs. 1(c) and 1(d), respectively. Complementary structural characterization of the as-grown and detached FeRh films can be found in the Supplemental Material, S1 [35].

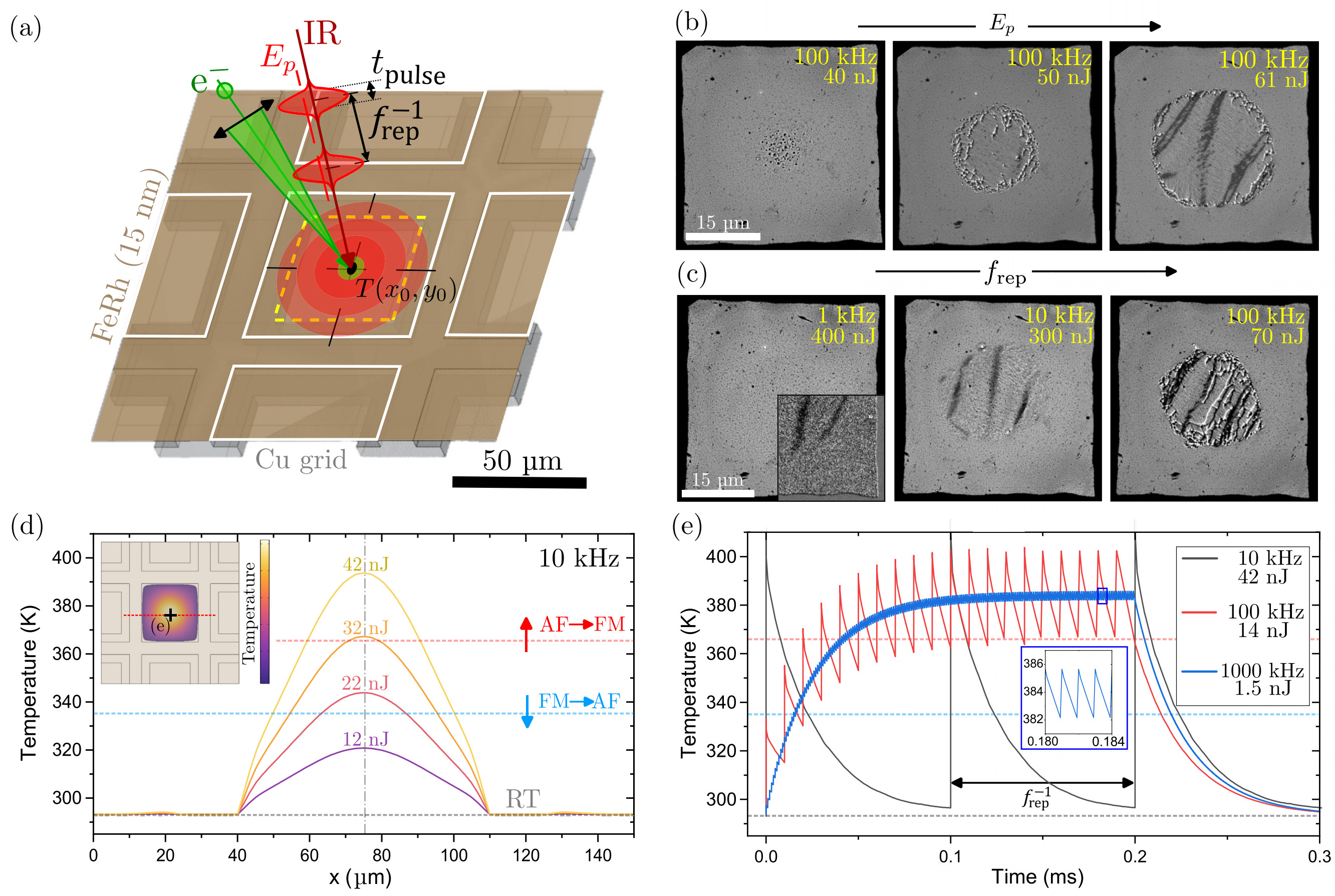

FIG. 2. Laser-induced FOPT in freestanding FeRh. (a) Scheme of the experimental configuration for laser-induced FOPT in TEM highlighting the repetition rate $f_{rep}$, laser pulse duration $t_{pulse}$, and energy per pulse $E_p$. (b), (c) LTEM images of the emerging FM phase for different $E_p$ and $f_{rep}$, respectively. (d) Simulated spatial distribution of the laser-induced peak temperature in FeRh (see the line profile in the inset) for different $E_p$ values. (e) Simulated time-dependent temperature of FeRh at the center of the laser-illuminated area for the $f_{rep}$ and $E_p$ values indicated in the legend, showing the effect of cumulative laser heating for higher repetition rates.

The thermally driven FOPT of FeRh can be revealed by *in situ* LTEM imaging, which is operated at large defocus (>1 mm) to visualize the walls of the FM domains. A series of LTEM images during the heating and cooling cycles are shown in Figs. 1(e) and 1(f). The formation of phases can be identified by the appearance of FM domain walls and AF/FM phase boundaries, which are visible as black and white lines in the LTEM images (see the Supplemental Material, Movies 1 and 2 [35]).

In order to evaluate the magnetic order parameter across the thermally driven FOPT, we quantify the relative content of the FM ($V_{FM}$) and AF regions ($V_{AF} = 1 - V_{FM}$) by estimating the area enclosed by sharp domain boundaries in the LTEM image, as indicated in Fig. 1(g). Estimation of the FM phase fraction across the entire series of temperature-dependent LTEM images allows reconstructing the local thermal hysteresis associated with the FOPT from a 5×5 $\mu m^2$ region [see Fig. 1(h)]. As the compressive strain imposed by the MgO substrate considerably relaxes in the freestanding FeRh film, we observe a significant drop in its phase transition temperature compared to the film supported on the substrate [see Fig. 1(h)], in accordance with the literature [36–38].

## III. LASER-DRIVEN FM PHASE NUCLEATION

To investigate the nucleation dynamics of the FOPT in FeRh, we employed *in situ* pulsed laser illumination inside the TEM. A comprehensive overview of the experimental setup is presented in Fig. 2(a) (see, also, Appendix A). The suspended FeRh thin film, supported on a Cu grid, was simultaneously exposed to a continuous electron beam and femtosecond laser pulses with a 1030-nm wavelength. The laser parameters, including the pulse duration $t_{pulse}$, repetition rate $f_{rep}$, and pulse energy $E_p$, were systematically varied to trigger the AF-to-FM phase transition. The influence of $E_p$ and $f_{rep}$ on the nucleation and growth of the FM phase is illustrated in Figs. 2(b) and 2(c), respectively.

Figure 2(b) presents the effect of increasing pulse energy $E_p$ at a fixed repetition rate. As $E_p$ increases, the FM phase emerges at the center of the laser-illuminated region and subsequently expands concentrically. The newly formed FM regions exhibit varying degrees of magnetic domain fragmentation, resulting in distinct domain-wall configurations that are readily resolved by LTEM. In addition to the magnetic contrast, prominent black stripes corresponding to bending contours of the suspended film are observed. These arise from topographic corrugations associated with the lattice expansion of the emerging FM phase relative to the surrounding AF matrix. Despite the surface corrugation, magnetization in the laser-induced FM regions is expected to orient predominantly in the sample plane due to the large demagnetizing field inherent to thin films. The observation of FM domain reversal upon applying an in-plane magnetic field using the TEM objective lens further supports the in-plane magnetization orientation of the nucleated FM domains (see the Supplemental Material, S2 [35]).

The evolution of FM contrast at laser repetition rates of 1, 10, and 100 kHz is shown in Fig. 2(c). A pronounced enhancement in FM phase contrast is observed upon increasing $f_{\mathrm{rep}}$, indicating that higher repetition rates do not allow sufficient cooling between pulses for the FM-to-AF transition to occur, and hence maintain the system in the high-temperature FM phase during illumination. The inset in the left image of Fig. 2(c) displays a contrast-enhanced differential image for the 1 kHz case, highlighting the subtle onset of FM nucleation under these conditions. The increasing contrast levels at 1 and 10 kHz are associated with the higher reproducibility of FM domains and AF/FM phase boundaries. Additionally, no significant influence of the pulse duration $t_{\mathrm{pulse}}$ on the resulting magnetic domain distribution was observed when comparing the effect of 0.2- and 5-ps-long laser pulses. This finding suggests that ultrafast nonequilibrium processes occurring immediately after excitation do not directly govern the spatial characteristics of the nucleated FM domains, such as position, size, or magnetization orientation. This observation is consistent with the reported insensitivity of the structural ordering timescale in FeRh to the duration of the pump pulse [30].

To further elucidate these experimental findings, numerical thermal simulations were performed using comsol multiphysics [39] to evaluate the temperature excursions generated by pulsed laser exposure and the subsequent cooling rates related to heat diffusion from the FeRh film toward the support Cu grid. The details of the simulation parameters and boundary conditions are provided in the Supplemental Material, S3 [35].

Figure 2(d) displays the simulated spatial temperature profile immediately following a single laser pulse. The Cu grid, being an efficient heat sink, maintains the suspended FeRh film regions in the proximity of the Cu grid at lower temperatures. In contrast, the center of the illuminated area reaches temperatures exceeding the AF-FM transition threshold (dashed red line), promoting FM phase formation preferentially in these areas.

The temporal evolution of temperature at the grid center for three different repetition rates is shown in Fig. 2(e). At 10 kHz, the temperature transiently exceeds the AF-FM transition threshold after each pulse, and subsequently falls below the FM-AF threshold before the next pulse arrives, indicating minimal thermal accumulation over time. In contrast, at 100 and 1000 kHz, cumulative heating gradually raises the temperature until reaching a steady-state value. In this regime, the temperature remains consistently above the FM-AF phase transition threshold, stabilizing the FM phase.

These observations explain both the variations of LTEM contrast for different repetition rates and the concentric growth of the FM phase for increasing laser power. Associated thermal simulations clarify the role of laser properties on the repeatability of the magnetic phases upon repeated laser exposure. In the following, we discuss the nucleation characteristics of the emerging FM phase and the associated role of laser exposure.

## IV. PULSED-LASER EXPOSURE-MEDIATED PHASE NUCLEATION KINETICS

The effect of laser excitation on the phase transition reveals two distinct nucleation regimes: *homogeneous* and *heterogeneous* nucleation. These terms refer to fundamentally different mechanisms governing the emergence of the FM phase from the parent AF phase.

In the *homogeneous nucleation* regime, the FM phase emerges through the spontaneous formation of a limited number of nucleation centers, uniformly distributed across the sample. Subsequent phase transformation proceeds primarily through lateral growth and coalescence of these domains. This regime dominates under static heating or low-energy pulsed excitation, where the local microstructure remains unaltered. In contrast, *heterogeneous nucleation* occurs when a high density of nucleation centers emerges at specific, structurally favorable sites such as dislocations or other lattice defects. In this regime, nucleation dominates over lateral domain propagation, leading to a distinct domain morphology. We show that cumulative pulsed-laser exposure can trigger this switch in the nucleation character, provided the excitation parameters exceed a well-defined threshold.

Specifically, we observed that heterogeneous nucleation in our sample geometry emerged only when the pulse energy exceeds 500 nJ per pulse (23 mJ/cm$^2$ for a beam diameter of 75 μm), while the repetition rate remains below 10 kHz, allowing repeated transitions between the AF and FM phases, as inferred from the simulated 10 kHz temperature profile in Fig. 2(e). The need for a repeated phase transition has been reproduced in different regions of the sample, but a systematic study as a function of repetition rate, fluence, and exposure time is beyond the scope of this work.

We note that an excessive increase in laser power leads to a permanent structural change and persistent FM phase even at room temperature [22]. We have verified this, for instance, by increasing the pulse energy above 400 nJ at 100 kHz. The exact threshold is again defined by a combination of factors, such as the repetition rate and exposure time. Within the studied fluence range, no signs of ablation or surface damage were observed, consistent with the much higher fluence values needed to reach the ablation regime reported in Ref. [22].

Figure 3 demonstrates the transition from homogeneous to heterogeneous nucleation in FeRh films induced by cumulative laser exposure. Figure 3(a) shows temperature-dependent magnetic domain configurations during static heating of two states of the same sample: a pristine film (framed in blue) and a laser-irradiated region (framed in red), which was exposed by pulsed laser for 60 s at a 10 kHz repetition rate, with 615 nJ per pulse. In the pristine film, the FM phase nucleates homogeneously, with uniformly distributed domains. In contrast, the laser-irradiated region displays heterogeneous nucleation, marked by a significantly higher density of nucleation centers. These nucleation sites are concentrated in the central part of the grid, which corresponds to the area most significantly thermally affected, as confirmed by thermal simulations. Figure 3(b) presents an analogous comparison under pulsed laser excitation, showing domain formation in the same two sample regions. At increasing pulse energies, the pristine film maintains homogeneous nucleation behavior, while the laser-treated area reveals localized, preferential nucleation sites, reaffirming the change in the nucleation regime under dynamic excitation conditions. Figure 3(c) quantifies the thermal requirements of the two nucleation regimes by plotting hysteresis curves reconstructed from the marked regions in Fig. 3(a). The onset of heterogeneous nucleation occurs nearly 20 K lower than in the homogeneous case, as marked by the red and blue filled dots. Furthermore, the laser-irradiated region shows an increase in hysteresis loop width by more than 5 K, reflecting an altered domain formation process.

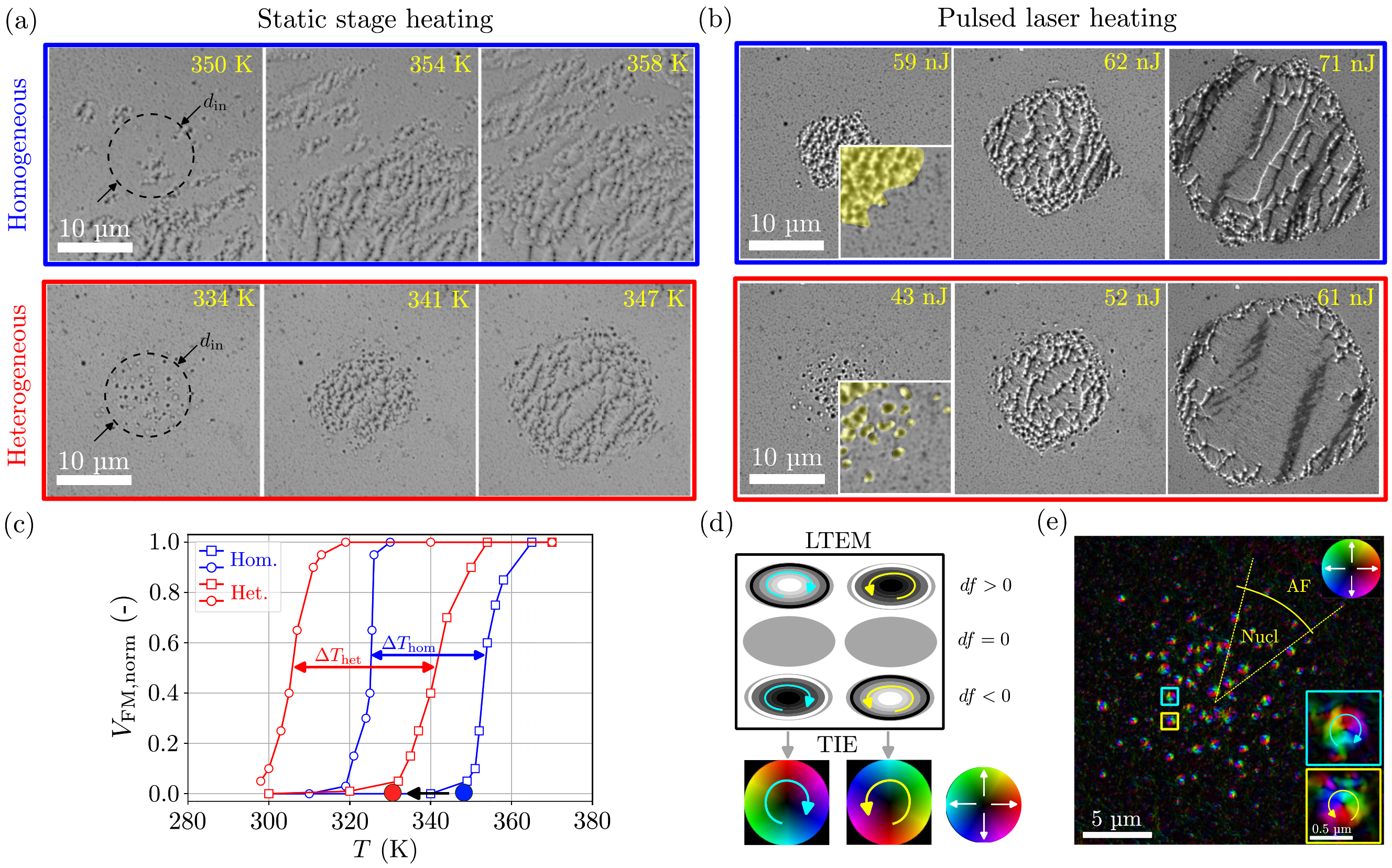

FIG. 3. Phase transition nucleation types in FeRh upon static and pulsed-laser heating. (a) Static heating dependence of LTEM contrast in the pristine film reveals a domain-wall propagation-dominated FOPT process (homogeneous, marked in blue). Imaging of the film regions irradiated above the specified threshold (see main text) reveals a nucleation-dominated FM phase formation (heterogeneous, marked in red). (b) Low-fluence laser-induced phase transition in the pristine film at a given energy per pulse shows a homogeneous nucleation type. In analogy to static heating, the laser-irradiated film reveals heterogeneous nucleation with lower nucleation thresholds and distinct domain morphology. In the homogeneous case, the FM phase forms as a single continuous domain (yellow area in the inset), whereas in the heterogeneous case, multiple isolated FM islands emerge at distinct nucleation sites, each marked by a yellow spot. (c) FM fraction hysteresis curves reconstructed from static heating, comparing homogeneous and heterogeneous nucleation. The zero value of FM phase content is defined by the lack of visibility of FM domains within the resolution limit of LTEM. Blue and red filled dots indicate the thermal offset of homogeneous and heterogeneous nucleation from 350 to 330 K, respectively. $\Delta T_{hom}$ and $\Delta T_{het}$ indicate the hysteresis widths of the transition for homogeneous and heterogeneous nucleation, respectively. The temperature-dependent FM fraction is calculated for the central area marked by a dashed circle in (a), with a diameter $d_{in} = 12$ μm. (d) TIE-based reconstruction scheme from LTEM figures for varying defocus values. (e) The in-plane magnetization vector map of the laser-induced heterogeneous FM nucleation reveals the existence of abundant magnetic vortices. The inset in (e) displays a zoomed-in view of two vortices with opposite circulations.

To investigate the magnetization textures that arise during heterogeneous nucleation, we take advantage of the transfer-of-intensity equation (TIE)-based reconstruction, which provides a vector-resolved map of in-plane magnetic induction. Figure 3(d) schematically indicates the reconstruction procedure from individual LTEM images taken at different defocus values. The resulting TIE-reconstructed image in Fig. 3(e) reveals a dense distribution of submicrometer magnetic vortices concentrated in the central film region. Their vortex circulation is reproducible across repeated experiments and remains unchanged under varying the pump laser polarization (see the Supplemental Material, S4 [35]), indicating that their magnetic configuration is governed by the local microstructure of the FeRh sample. As the laser fluence increases, new vortices are nucleated until they coalesce into a continuous FM region.

Figure 4 presents a quantitative analysis of the two nucleation regimes induced by low-fluence pulsed laser excitation.

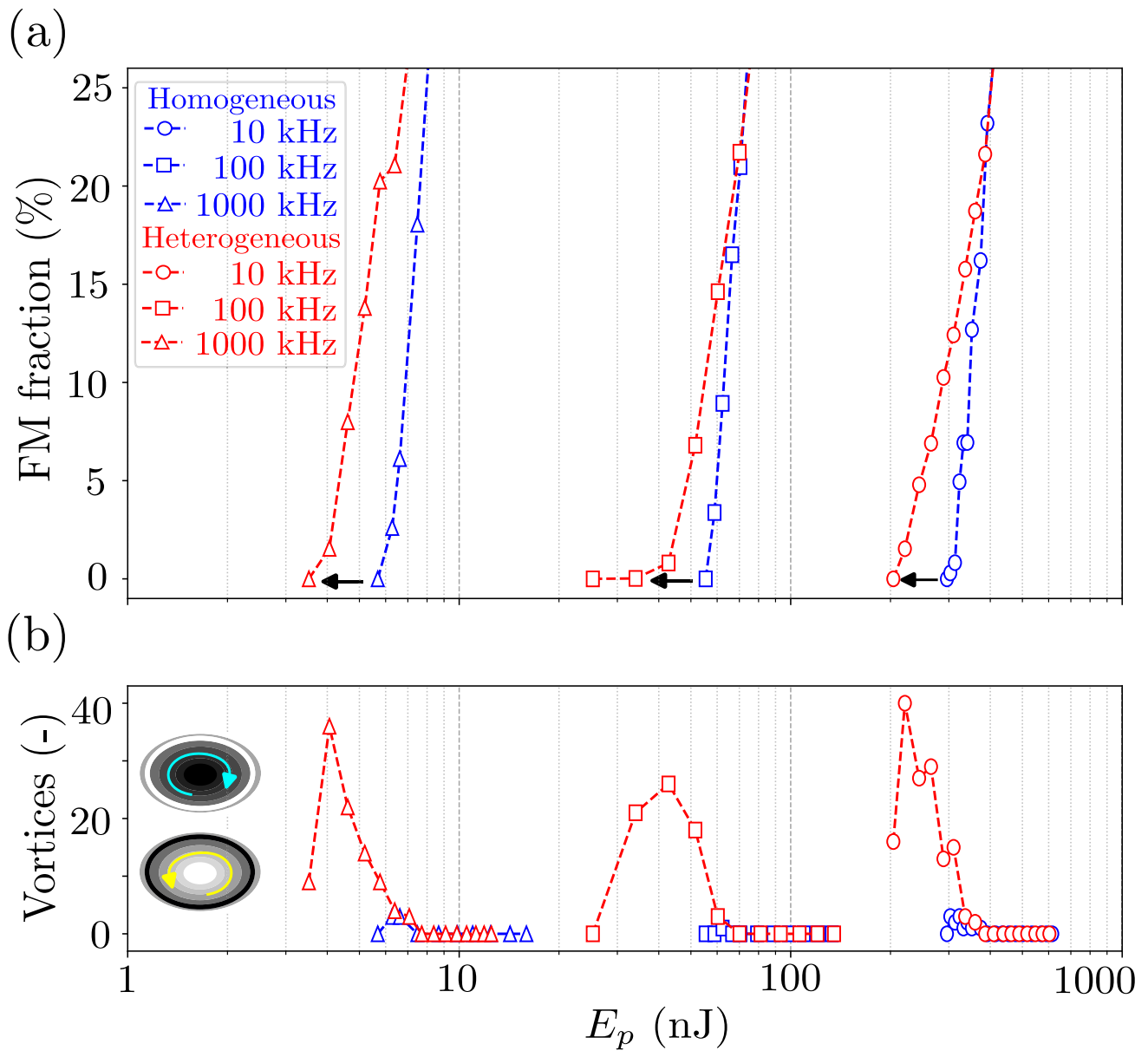

FIG. 4. Magnetic configurations upon homogeneous and heterogeneous nucleation. (a) FM fraction as a function of laser pulse energy at different repetition rates, revealing lower nucleation thresholds for heterogeneous nucleation in the inner region compared to the homogeneous nucleation regime. (b) Number of nucleated magnetic vortices as a function of pulse energy, showing their preferential formation in the heterogeneous nucleation regime.

Figure 4(a) shows the fraction of the FM phase as a function of pulse energy $E_p$ for various laser repetition rates. A significant reduction in the nucleation threshold is observed within the laser-processed region, reaching up to a 50% decrease in the required pulse energy, particularly at higher repetition rates. This confirms that the cumulative laser action facilitates FM phase formation under less energetically demanding conditions. Figure 4(b) quantifies the number of isolated FM vortices corresponding to the same experimental conditions. A marked increase in vortex density is observed under heterogeneous nucleation conditions, indicating that the FM phase forms via fragmentation into multiple topological domains. In contrast, the homogeneous regime shows few or no vortices, which is consistent with smoother, propagation-driven domain growth.

The appearance of magnetic vortices as a recurring motif in the heterogeneous regime aligns with previous observations of flux-closure structures in FeRh thin films. Baldasseroni *et al.* imaged such configurations using x-ray magnetic microscopy across the thermally driven FOPT [40], attributing them to the inhomogeneous phase nucleation caused by crystallographic grain boundaries. Similarly, Almeida *et al.* also observed a few instances of vortexlike magnetic configurations in freestanding FeRh films via TEM [41], suggesting a connection to strain relaxation and mechanical boundary conditions. Notably, we do not observe nucleation through uniformly magnetized FM domains, which have been reported in prior studies [40]. Their absence in our experiments may stem from the reduced film thickness and the continuous imaging of the laser-induced phase transition. The uniform domains, lacking topological protection, are more vulnerable to fluctuations induced by repeated optical stimulation and thus more difficult to capture by continuous *in situ* imaging. In contrast, magnetic vortices are stabilized by their topology and thus their contrast persists over repeated laser cycles.

Together, these results demonstrate that cumulative femtosecond laser exposure not only reduces the thermal energy required for nucleation, but also fundamentally alters the dominant nucleation mechanism. This shift is directly reflected in the topology and morphology of the emergent FM textures.

## V. MICROSCOPIC ORIGIN OF HETEROGENEOUS NUCLEATION IN FeRh FILMS

To uncover the microscopic mechanism behind the heterogeneous laser-induced phase nucleation in FeRh, we performed combined *in situ* LTEM imaging and high-resolution structural characterization on the freestanding film exposed to cumulative pulsed laser excitation.

Figure 5(a) shows the onset of FM domain formation around a localized, high-contrast structural feature with an approximate size of a few tens of nanometers (marked in red). This object acts as a nucleation center from which FM regions grow upon repeated laser pulses. The corresponding TIE-reconstructed magnetic induction map in Fig. 5(b) reveals a vortexlike magnetic texture centered on this nucleation seed. In addition, a secondary population of smaller vortexlike structures with diameters around 100 nm appears across the field of view. However, only the larger vortices are observed to facilitate domain growth during subsequent laser exposure, while the smaller features are likely associated with phase contrast from surface crystallites or stoichiometric inhomogeneities (see the Supplemental Material, S5 [35]). These features also contribute to the overall LTEM contrast, as seen, for instance, in the form of the dark spots in Figs. 1(e) and 1(f).

To investigate the origin of the nucleation centers in the heterogeneous regime, we examined their local structural environment. Figure 5(c) presents a TEM image acquired under the two-beam diffraction condition, highlighting strain fields associated with a dense dislocation network. These strain-accumulated regions feature a pronounced dark contrast in TEM and are strongly correlated with magnetic vortex nucleation centers. HR-STEM imaging in Fig. 5(d) further resolves the dislocation types: the left panel shows an edge dislocation with Burgers vector $b = [001]$, while the right panel shows a screw dislocation with $b = [111]$, confirming altogether the formation of a mixed-character dislocation network under repeated pulsed-laser excitation. Figure 5(e) summarizes these findings schematically, illustrating how local dislocation networks and associated strain fields make vortexlike magnetic textures preferential for FM phase nucleation.

These results provide direct evidence that structural defects, particularly highly strained, dislocation-rich regions, serve as efficient nucleation centers in FeRh. This conclusion interlinks the previously observed strain-mediated phase transition in FeRh [38,42] with dislocation networks mediating the phase transitions [1]. Additionally, being able to visualize the individual nucleation centers offers insight into the balance between nucleation and domain propagation in the overall phase transformation process. Recent studies have shown that preexisting FM nuclei can strongly influence subsequent laser-induced transitions [43], highlighting the potential of coherent laser excitation to control the density and position of nucleation centers. Our results establish a microscopic foundation for such control strategies by linking the individual nucleation centers to local structural modifications induced by laser exposure.

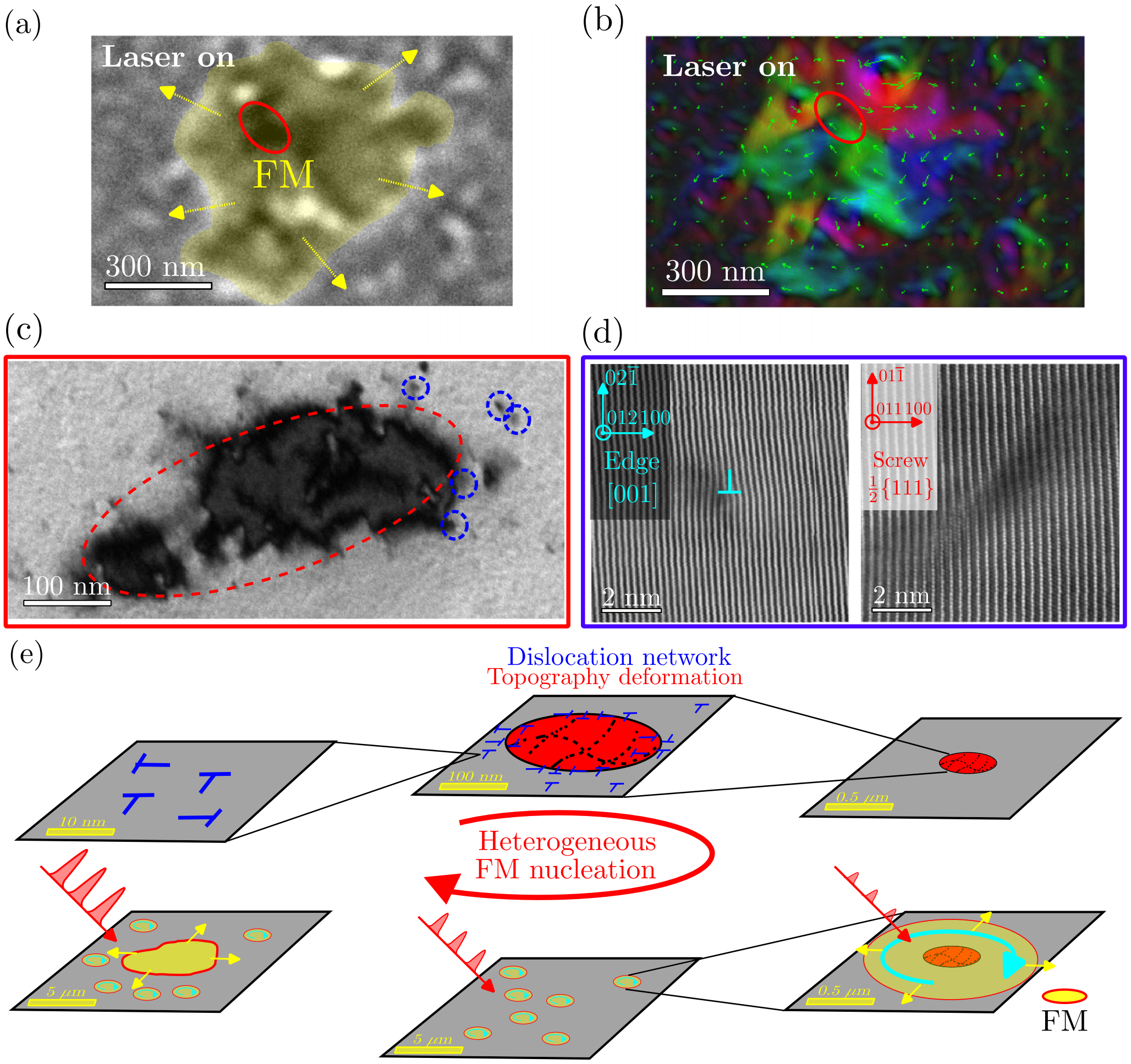

FIG. 5. Structural origin of the heterogeneous nucleation of FM phase in FeRh. (a) LTEM figure showing the radial expansion of the vortex from a dark-contrast seed upon laser exposure. (b) TIE-based reconstruction of the magnetic texture from (a), revealing details of vortexlike magnetic contrast at the nucleation site. (c) Magnified bright-field TEM image under the two-beam diffraction condition. The dark contrast reveals a strain-accumulated region associated with a dense network of dislocations, which promotes FM nucleation. (d) HR-STEM images of an edge dislocation with Burgers vector $b = [001]$ (left panel) and a screw dislocation with $b = [111]$ (right panel). (e) Schematic overview of the proposed dislocation-driven nucleation mechanism: laser-induced dislocation networks and topographic deformation generate favorable conditions for heterogeneous nucleation, with magnetic vortices being preferred for FM configurations.

## VI. DISCUSSION

The *in situ* LTEM investigation of freestanding FeRh films under pulsed-laser excitation provides detailed insights into the local nucleation mechanisms of the magnetostructural phase transition. Our results reveal a profound alteration of the FM phase nucleation pathway from a homogeneous to a heterogeneous regime, which is driven by the cumulative effect of laser exposure on the defect landscape within the material. Magnetic vortices are identified as the dominant nucleation texture in the heterogeneous regime. These vortices consistently form in strain-rich regions and act as preferential nucleation centers, indicating a link between topological spin textures, local strain, and defect-stabilized nucleation behavior. We observe a decrease in the transition temperature from 350 K upon homogeneous nucleation to 330 K upon heterogeneous nucleation, with a corresponding 50% reduction in the laser fluence needed to trigger the phase transition.

It was observed that the modification of the phase transition nucleation mechanism is closely associated with defect formation upon laser exposure, which is primarily driven by a combination of high strain, rapid thermal expansion, and fast quenching. The 1% lattice volume change accompanying the transition, combined with spatial gradients of the laser-induced fast thermal expansion (reaching up to 0.5% across the HWHM of the beam spot for moderate fluences [29]), generates substantial internal stresses. Using Eshelby's inclusion model, the hydrostatic stress associated with the formation of a FM nucleus within the AF matrix can be estimated to reach approximately −2.3 GPa (see quantification in Appendix B), arising from a total volumetric strain of $\varepsilon_V \approx 0.010\,75$ that combines transformation and thermal mismatch components.

Even allowing for partial strain relaxation and nonspherical domain shapes, the resulting local stresses remain of the order of gigapascals—sufficient to induce significant defect motion and reconfiguration through the Peach-Koehler mechanism [44]. Extreme thermal gradients and temperature ramp rates up to $10^{14}$ K/s during heating and $10^7$–$10^9$ K/s during cooling further contribute to defect stabilization through quenching. Experimentally, the shift in the hysteresis loop to lower temperatures and a slight broadening after laser exposure may indicate an increased defect density, which promotes nucleation of the FM phase at lower temperatures, as shown in Fig. 3(c). Direct structural observations before and after laser exposure, presented in the Supplemental Material, S6 [35], confirm the formation and reorganization of dislocations in the laser-illuminated regions.

Under high-repetition-rate laser irradiation, the repeated AF-FM transformations subject the film to billions of stress cycles in the gigapascal range, approaching the yield stress of FeRh. In a substrate-clamped geometry, these stresses would be largely accommodated elastically, but in a freestanding film their full amplitude is realized, enabling microstructural reconfiguration through repeated transformational stress excursions. Although the stresses are insufficient to trigger new dislocation multiplication, they readily mobilize preexisting misfit dislocations, which can reorganize into stable, locked (Frank) networks that locally relax compressive stress. Upon cooling to the AF phase, these strain-relieved regions retain residual tensile stress and subsequently act as preferential nucleation centers for FM vortices during the next transformation cycle.

Notably, the formation of a Frank network of defects [45] results in cumulative strain fields reaching up to 200 nm in size that form highly strained and locally wrinkled regions in the freestanding film, which are evidenced by high diffraction contrast, as shown in Fig. 5(c). These regions stabilize the FM phase and act as preferential nucleation centers for the FM phase.

Previous studies have linked dislocations to chiral magnetic textures through defect-induced Dzyaloshinskii-Moriya–type interactions, which stabilize helical or skyrmionic states [46,47]. In the present case, however, the connection between dislocations and magnetic vortices is interpreted purely in terms of magnetostatics. The agglomerated dislocations act as localized strain centers that lower the energy barrier for the AF $\rightarrow$FM transformation, promoting the nucleation of thin FM islands whose lateral extent is defined by the strain field. Within such confined FM regions, minimization of magnetostatic energy naturally favors flux-closure or vortexlike domain configurations [48,49], establishing a direct link between the defect structure, strain, and magnetostatically promoted vortex formation in FM islands of FeRh.

The cumulative modification of nucleation properties observed under repeated laser exposure raises critical questions for ultrafast pump-probe experiments, where long-term laser exposure may inadvertently alter the energy landscape of the material. These findings underscore the necessity of carefully controlling laser fluence and defect formation in future time-resolved investigations of phase transitions in magnetic materials. The broader implications of defect-mediated nucleation extend beyond the current system, offering new avenues for tailoring phase transitions in functional materials via controlled defect engineering. The demonstrated sensitivity of nucleation pathways to photo-induced structural modifications suggests that nucleation properties themselves may be tunable using ultrafast laser pulses, opening new directions for phase transition control via photoexcitation.

## VII. CONCLUSION

We demonstrate using *in situ* TEM that pulsed-laser exposure of freestanding FeRh thin films significantly modifies the magnetostructural phase transition. It introduces local defect networks, which fundamentally alter the nucleation pathway from homogeneous to heterogeneous. The defect-mediated nucleation regime reduces the laser fluence threshold by up to 50% and promotes the emergence of magnetic vortices as preferential nucleation motifs. These findings underscore the critical role of defect formation in tailoring phase transition properties and offer insights into the interplay between laser-induced strain, defect stabilization, and transformation dynamics in magnetic materials.

In addition, the possibility to purposefully generate and control magnetic textures, such as vortices, during a phase transition opens exciting opportunities for applications in reconfigurable magnetic memory, ultrafast spintronic devices, and topological information storage. This work demonstrates a pathway for defect- and texture-engineered phase transitions, with direct implications for ultrafast stroboscopic experiments and the design of functional materials with tunable nucleation landscapes.

## ACKNOWLEDGMENTS

We thank Dr. Jakub Zlámal for assistance with Comsol simulations. J.H. was supported by the Thermo Fisher Scientific scholarship and the CEITEC internal grant agency. This work was supported by the TERAFIT Project No. CZ.02.01.01/00/22_008/0004594 funded by OP JAK, call Excellent Research. Access to the CEITEC Nano Research Infrastructure was supported by the Ministry of Education, Youth and Sports (MEYS) of the Czech Republic under the CzechNanoLab (Project No. LM2023051). A.A., P.C., F.C., and T.L. acknowledge the support by the Air Force Office of Scientific Research under Award No. FA8655-24-1-7015.

## DATA AVAILABILITY

The data that support the findings of this article are openly available [50]; embargo periods may apply.

## APPENDIX A: EXPERIMENTAL METHODS

### 1. FeRh film growth and characterization

Epitaxial FeRh films were deposited using a magnetron sputtering system (BESTEC) with a base pressure lower than $5\times10^{-8}$ mbar. 15-nm-thick films were sputtered from an equiatomic FeRh target on single-crystal MgO(001) substrates. The substrates were preheated to 723 K for 1 h before deposition. Film growth was carried out at the same temperature at an Ar pressure of $2.7\times10^{-3}$ mbar. A sputter power of 50 W resulted in a deposition rate of 0.3 Å/s. The films were postgrowth annealed at 1000 K for 30 min to obtain the desired B2 crystallographic structure. The samples were then cooled in high vacuum and capped with a 3-nm-thick Al layer once at room temperature.

The film structure was analyzed by x-ray reflectivity (XRR) and diffraction (XRD) employing a Rigaku SmartLab (9 kW) diffractometer with Cu $K_\alpha$ radiation ($\lambda = 1.5406$ Å).

The temperature dependence of magnetization was measured via vibrating sample magnetometry (VSM) under an in-plane magnetic field of 3 T using a Quantum Design VersaLab magnetometer. The field-induced shift of the phase transition temperature (corresponding to $-8$ K/T [16]) was compensated for the shown VSM datasets. All magnetization data are presented after subtracting the diamagnetic substrate contribution.

#### 2. Freestanding FeRh films

FeRh thin films were detached from the substrate by chemically dissolving MgO in an aqueous solution of the disodium salt of ethylenediaminetetraacetic acid (EDTA), following the procedures described in Refs. [32–34]. The FeRh film on the MgO substrate was initially cut with a diamond saw into $2\times1$ mm$^2$ pieces, which were immersed in an aqueous EDTA solution (molar concentration 0.3 M). The etching lasted for 10 h at 348 K, after which the released film was transferred to a Cu TEM grid and gently washed in deionized water to remove salt residuals.

#### 3. Transmission electron microscopy

*Magnetic imaging in TEM*. A JEOL JEM-2100 TEM, operated at 200 keV with a thermionic $LaB_6$ gun customized for a femtosecond laser pump-probe setup [51], was employed in the experiments. The microscope is equipped with a Gatan GIF Quantum 895 energy filter and a US1000 CCD camera used for all image acquisitions. The sample was loaded in a single-tilt Gatan Cryostage 626, allowing heating up to 385 K. LTEM images were acquired in a dedicated *Low Mag* mode of the JEOL JEM-2100, with the objective lens set to 0.0 excitation. The remanent magnetic field in the sample plane is estimated to be below 5 mT. Reconstruction using the transfer of intensity equation (TIE) was performed using the hrem qpt module version 4.0.2 [52] on a symmetrical TIE dataset (in-focus $df = 0$, overfocus $df = +D$, underfocus $df = -D$) that was manually aligned and phase reconstructed (high pass filter applied), and from which the magnetic field component was obtained.

*Ultrafast laser setup*. An Amplitude SAS (Satsuma) fiber laser operated at 1030 nm (maximum output power 20 W, minimum pulse duration 200 fs) was coupled to the TEM setup. The sample is illuminated by focusing the light to a spot of about 75 μm in the FWHM of the beam intensity profile (using a lens with a focal length of 300 mm). The laser's repetition rate $f_{rep}$ was set to 1 MHz, while an external modulator allowed varying it in the 1–100 kHz range. The pulse length was adjusted by the position of a sapphire crystal compressor to be either 0.2 or 5.0 ps. The incident laser power was modified with the aid of a $\lambda/2$ waveplate and a thin-film polarizer. The energy per pulse was determined using a power meter, and the laser spot's spatial profile was monitored by a profiler.

*SAED data*. The diffraction patterns were acquired in parallel beam mode with a 2 μm selected area aperture and using a 100 mm camera length of the projection system.

*Structural TEM analysis*. High-resolution structural TEM analysis was performed using a FEI Titan Themis instrument operated at 300 kV and equipped with an X-FEG emitter and double aberration correction.

*Chemical TEM analysis*. Composition analysis of the laser-exposed areas of FeRh was performed using STEM Energy Dispersive X-ray Spectroscopy (EDX) in a Thermo Fisher Scientific Talos F200S microscope operated at 200 keV and equipped with two in-column Super-X EDX detectors.

## APPENDIX B: STRESS QUANTIFICATION UPON LASER PULSING

The hydrostatic stress ($\sigma_h$) associated with nucleating a FM domain within the AF matrix of FeRh can be estimated using Eshelby's solution for an ellipsoidal inclusion embedded in an identical matrix [53],

$$\sigma_h = \frac{12KG}{3K + 4G}\varepsilon_V. \tag{B1}$$

In this framework, the FM nucleus is treated as a coherent spherical inclusion that experiences an isotropic volumetric eigenstrain. This eigenstrain ($\varepsilon_V$) has two main contributions,

$$\varepsilon_V = \varepsilon_{Tr} + \varepsilon_{Th}. \tag{B2}$$

The transformation strain ($\varepsilon_{Tr}$) arises from the AF-FM transition, which is about 1% volume expansion, and a thermal mismatch strain ($\varepsilon_{Th}$) is given by

$$\varepsilon_{Th} = 3(\alpha_{FM} - \alpha_{AF})\Delta T. \tag{B3}$$

Here, $\alpha_{FM}$ and $\alpha_{AF}$ are the thermal expansion coefficients for FeRh individual magnetic phases, and $\Delta T$ is the temperature change across the transition. For FeRh, the bulk modulus is approximately 197 GPa, and the shear modulus, assuming a Poisson's ratio of 0.31, is about 86 GPa [54–57]. With a 1% transformation strain, the stress is of the order of $-2.2$ GPa, and including a typical thermal mismatch strain of about $7.5\times10^{-4}$ for $\Delta\alpha \approx 5\times10^{-6}$ K$^{-1}$ over a 50 K temperature change increases the magnitude of approximately $-2.3$ GPa. This is considerable stress, and the bend contours observed in the FM phase clearly indicate its magnitude, deforming and buckling the laser-induced FM phase in the film. Due to differences in elastic moduli between phases, nonspherical domain shapes, partial strain relaxation, and the presence of defects, the actual value will be reduced by a factor of 2 or more.

[1] X. Zhang, J. Zhang, H. Wang, J. Rogal, H.-Y. Li, S.-H. Wei, and T. Hickel, Defect-characterized phase transition kinetics, Appl. Phys. Rev. **9**, 041311 (2022).

[2] W. Zhang, A. Thiess, P. Zalden, R. Zeller, P. H. Dederichs, J. Y. Rat, M. Wuttig, S. Blugel, and R. Mazzarello, Role of vacancies in metal-insulator transitions of crystalline phase-change materials, Nat. Mater. **11**, 952 (2012).

[3] M. Wuttig and N. Yamada, Phase-change materials for rewriteable data storage, Nat. Mater. **6**, 824 (2007).

[4] S. W. Nam, H. S. Chung, Y. C. Lo, L. Qi, J. Li, Y. Lu, A. T. Johnson, Y. Jung, P. Nukala, and R. Agarwal, Electrical wind force-driven and dislocation-templated amorphization in phase-change nanowires, Science **336**, 1561 (2012).

[5] G. Modi, E. A. Stach, and R. Agarwal, Low-power switching through disorder and carrier localization in bismuth-doped germanium telluride phase change memory nanowires, ACS Nano **14**, 2162 (2020).

[6] X. Zhou, J. Kalikka, X. Ji, L. Wu, Z. Song, and R. E. Simpson, Phase-Change memory materials by design: A strain engineering approach, Adv. Mater. **28**, 3007 (2016).

[7] A. S. McLeod, E. van Heumen, J. G. Ramirez, S. Wang, T. Saerbeck, S. Guenon, M. Goldflam, L. Anderegg, P. Kelly, A. Mueller, M. K. Liu, I. K. Schuller, and D. N. Basov, Nanotextured phase coexistence in the correlated insulator V2O3, Nat. Phys. **13**, 80 (2017).

[8] Y. Wang, X. Sun, Z. Chen, Z. Cai, H. Zhou, T.-M. Lu, and J. Shi, Defect-engineered epitaxial VO$(2 \pm \delta)$ in strain engineering of heterogeneous soft crystals, Sci. Adv. **4**, eaar3679 (2018).

[9] D. Song, L. Ma, S. Zhou, and J. Zhu, Oxygen deficiency induced deterioration in microstructure and magnetic properties at $Y_3Fe_5O_{12}$/Pt interface, Appl. Phys. Lett. **107**, 042401 (2015).

[10] A. Lak, S. Disch, and P. Bender, Embracing defects and disorder in magnetic nanoparticles, Adv. Sci. **8**, 2002682 (2021).

[11] Y. Cheng, A. Zong, L. Wu, Q. Meng, W. Xia, F. Qi, P. Zhu, X. Zou, T. Jiang, Y. Guo, J. van Wezel, A. Kogar, M. W. Zuerch, J. Zhang, Y. Zhu, and D. Xiang, Ultrafast formation of topological defects in a two-dimensional charge density wave, Nat. Phys. **20**, 54 (2024).

[12] A. S. Johnson, D. Perez-Salinas, K. M. Siddiqui, S. Kim, S. Choi, K. Volckaert, P. E. Majchrzak, S. Ulstrup, N. Agarwal, K. Hallman, R. F. Haglund, C. M. Günther, B. Pfau, S. Eisebitt, D. Backes, F. Maccherozzi, A. Fitzpatrick, S. S. Dhesi, P. Gargiani, M. Valvidares, *et al.*, Ultrafast x-ray imaging of the light-induced phase transition in $VO_2$, Nat. Phys. **19**, 215 (2023).

[13] A. Zong, A. Kogar, Y.-Q. Bie, T. Rohwer, C. Lee, E. Baldini, E. Ergeçen, M. B. Yilmaz, B. Freelon, E. J. Sie, H. Zhou, J. Straquadine, P. Walmsley, P. E. Dolgirev, A. V. Rozhkov, I. R. Fisher, P. Jarillo-Herrero, B. V. Fine, and N. Gedik, Evidence for topological defects in a photoinduced phase transition, Nat. Phys. **15**, 27 (2019).

[14] P. Olleros-Rodríguez, M. Strungaru, S. Ruta, P. I. Gavriloaea, A. Gudín, P. Perna, R. Chantrell, and O. Chubykalo-Fesenko, Non-equilibrium heating path for the laser-induced nucleation of metastable skyrmion lattices, Nanoscale **14**, 15701 (2022).

[15] J. S. Kouvel and C. C. Hartelius, Anomalous magnetic moments and transformations in the ordered alloy FeRh, J. Appl. Phys. **33**, 1343 (1962).

[16] S. Maat, J. U. Thiele, and E. E. Fullerton, Temperature and field hysteresis of the antiferromagnetic-to-ferromagnetic phase transition in epitaxial FeRh films, Phys. Rev. B **72**, 214432 (2005).

[17] D. W. Cooke, F. Hellman, C. Baldasseroni, C. Bordel, S. Moyerman, and E. E. Fullerton, Thermodynamic measurements of Fe-Rh alloys, Phys. Rev. Lett. **109**, 255901 (2012).

[18] L. H. Lewis, C. H. Marrows, and S. Langridge, Coupled magnetic, structural, and electronic phase transitions in FeRh, J. Phys. D **49**, 323002 (2016).

[19] K. Aikoh, S. Kosugi, T. Matsui, and A. Iwase, Quantitative control of magnetic ordering in FeRh thin films using 30 keV Ga ion irradiation from a focused ion beam system, J. Appl. Phys. **109**, 07E311 (2011).

[20] D. G. Merkel, A. Lengyel, D. L. Nagy, A. Németh, Z. E. Horváth, C. Bogdán, M. A. Gracheva, G. Hegedűs, S. Sajti, G. Z. Radnóczi, and E. Szilágyi, Reversible control of magnetism in FeRh thin films, Sci. Rep. **10**, 13923 (2020).

[21] S. Song, C.-W. Cho, J. Kim, J. Lee, D. Lee, D. Kim, H. Kim, H. Kang, C.-H. Park, J. K. Park, J. H. Jang, and S. Park, Correlation between phase transition characteristics and hydrogen irradiation-induced Frenkel defect formations in FeRh films, J. Alloys Compd. **901**, 163611 (2022).

[22] D. G. Merkel, K. Sájerman, T. Váczi, S. Lenk, G. Hegedűs, S. Sajti, A. Németh, M. A. Gracheva, P. Petrik, D. Mukherjee, Z. E. Horváth, D. L. Nagy, and A. Lengyel, Laser irradiation effects in FeRh thin film, Mater. Res. Express **10**, 076101 (2023).

[23] K. Aikoh, A. Tohki, T. Matsui, A. Iwase, T. Satoh, K. Takano, M. Kohka, Y. Saitoh, T. Kamiya, T. Ohkochi, M. Kotsugi, T. Nakamura, and T. Kinoshita, MFM and PEEM observation of micrometre-sized magnetic dot arrays fabricated by ion-microbeam irradiation in FeRh thin films, J. Synchrotron Radiat. **19**, 223 (2012).

[24] G. Ju, J. Hohlfeld, B. Bergman, R. J. M. van de Veerdonk, O. N. Mryasov, J.-Y. Kim, X. Wu, D. Weller, and B. Koopmans, Ultrafast generation of ferromagnetic order via a laser-induced phase transformation in FeRh thin films, Phys. Rev. Lett. **93**, 197403 (2004).

[25] S. O. Mariager, F. Pressacco, G. Ingold, A. Caviezel, E. Möhr-Vorobeva, P. Beaud, S. L. Johnson, C. J. Milne, E. Mancini, S. Moyerman, E. E. Fullerton, R. Feidenhansl, C. H. Back, and C. Quitmann, Structural and magnetic dynamics of a laser induced phase transition in FeRh, Phys. Rev. Lett. **108**, 087201 (2012).

[26] F. Pressacco, D. Sangalli, V. Uhlíř, D. Kutnyakhov, J. A. Arregi, S. Y. Agustsson, G. Brenner, H. Redlin, M. Heber, D. Vasilyev, J. Demsar, G. Schönhense, M. Gatti, A. Marini, W. Wurth, and F. Sirotti, Subpicosecond metamagnetic phase transition in FeRh driven by nonequilibrium electron dynamics, Nat. Commun. **12**, 5088 (2021).

[27] K. Kang, H. Omura, D. Yesudas, O. Lee, K.-J. Lee, H.-W. Lee, T. Taniyama, and G.-M. Choi, Spin current driven by ultrafast magnetization of FeRh, Nat. Commun. **14**, 3619 (2023).

[28] D. Hamara, M. Strungaru, J. R. Massey, Q. Remy, X. Chen, G. Nava Antonio, O. Alves Santos, M. Hehn, R. F. Evans, R. W. Chantrell, *et al.*, Ultra-high spin emission from antiferromagnetic FeRh, Nat. Commun. **15**, 4958 (2024).

[29] M. Mattern, J. Jarecki, J. A. Arregi, V. Uhlíř, M. Rössle, and M. Bargheer, Speed limits of the laser-induced phase transition in FeRh, APL Mater. **12**, 051124 (2024).

[30] M. Mattern, S. P. Zeuschner, M. Rössle, J. A. Arregi, V. Uhlíř, and M. Bargheer, Non-thermal electrons open the nonequilib-

rium pathway of the phase transition in FeRh, Commun. Phys. **8**, 140 (2025).

[31] I. A. Dolgikh, T. G. H. Blank, A. G. Buzdakov, G. Li, K. H. Prabhakara, S. K. K. Patel, R. Medapalli, E. E. Fullerton, O. V. Koplak, J. H. Mentink, K. A. Zvezdin, A. K. Zvezdin, P. C. M. Christianen, and A. V. Kimel, Ultrafast emergence of ferromagnetism in antiferromagnetic FeRh in high magnetic fields, npj Spintron. **3**, 5 (2025).

[32] T. Edler and S. G. Mayr, Film lift-off from MgO: Freestanding single crystalline Fe-Pd films suitable for magnetic shape memory actuation - and beyond, Adv. Mater. **22**, 4969 (2010).

[33] L. Motyčková, J. A. Arregi, M. Staňo, S. Průša, K. Částková, and V. Uhlíř, Preserving metamagnetism in self-assembled ferh nanomagnets, ACS Appl. Mater. Interfaces **15**, 8653 (2023).

[34] S. Gorji, A. Kashiwar, L. S. Mantha, R. Kruk, R. Witte, P. Marek, H. Hahn, C. Kübel, and T. Scherer, Nanowire facilitated transfer of sensitive TEM samples in a FIB, Ultramicroscopy **219**, 113075 (2020).

[35] See Supplemental Material at http://link.aps.org/supplemental/10.1103/k8nj-p557 for additional structural characterization of the FeRh films, the role of the TEM objective field on FM domains, details of thermal simulations, the role of the laser pump polarization, the stoichiometry analysis of freestanding FeRh films after strong laser illumination, and weak-beam imaging of dislocations in FeRh films, which includes Refs. [58–61].

[36] Y. Ohtani and I. Hatakeyama, Features of broad magnetic transition in FeRh thin film, J. Magn. Magn. Mater. **131**, 339 (1994).

[37] A. Ceballos, Z. Chen, O. Schneider, C. Bordel, L.-W. Wang, and F. Hellman, Effect of strain and thickness on the transition temperature of epitaxial FeRh thin-films, Appl. Phys. Lett. **111**, 172401 (2017).

[38] J. A. Arregi, O. Caha, and V. Uhlíř, Evolution of strain across the magnetostructural phase transition in epitaxial FeRh films on different substrates, Phys. Rev. B **101**, 174413 (2020).

[39] Computer software COMSOL Multiphysics® v. 6.2; http://www.Comsol.com (COMSOL AB, Stockholm, Sweden).

[40] C. Baldasseroni, C. Bordel, A. X. Gray, A. M. Kaiser, F. Kronast, J. Herrero-Albillos, C. M. Schneider, C. S. Fadley, and F. Hellman, Temperature-driven nucleation of ferromagnetic domains in FeRh thin films, Appl. Phys. Lett. **100**, 262401 (2012).

[41] T. P. Almeida, D. McGrouther, R. Temple, J. Massey, Y. Li, T. Moore, C. H. Marrows, and S. McVitie, Direct visualization of the magnetostructural phase transition in nanoscale FeRh thin films using differential phase contrast imaging, Phys. Rev. Mater. **4**, 034410 (2020).

[42] J. A. Arregi, F. Ringe, J. Hajduček, O. Gomonay, T. Molnár, J. Jaskowiec, and V. Uhlíř, Magnetic-field-controlled growth of magnetoelastic phase domains in FeRh, J. Phys.: Mater. **6**, 034003 (2023).

[43] G. Li, R. Medapalli, J. H. Mentink, R. V. Mikhaylovskiy, T. G. Blank, S. K. Patel, A. K. Zvezdin, T. Rasing, E. E. Fullerton, and A. V. Kimel, Ultrafast kinetics of the antiferromagnetic-ferromagnetic phase transition in FeRh, Nat. Commun. **13**, 2998 (2022).

[44] M. Peach and J. S. Koehler, The forces exerted on dislocations and the stress fields produced by them, Phys. Rev. **80**, 436 (1950).

[45] M. Delfani and F. Kakavand, Frank network of dislocations within Mindlin's second strain gradient theory of elasticity, Intl. J. Mech. Sci. **164**, 105150 (2019).

[46] A. Michels, D. Mettus, I. Titov, A. Malyeyev, M. Bersweiler, P. Bender, I. Peral, R. Birringer, Y. Quan, P. Hautle, J. Kohlbrecher, D. Honecker, J. R. Fernández, L. F. Barquín, and K. L. Metlov, Microstructural-defect-induced Dzyaloshinskii-Moriya interaction, Phys. Rev. B **99**, 014416 (2019).

[47] M. Azhar, V. P. Kravchuk, and M. Garst, Screw dislocations in chiral magnets, Phys. Rev. Lett. **128**, 157204 (2022).

[48] R. P. Cowburn, Property variation with shape in magnetic nanoelements, J. Phys. D **33**, R1 (2000).

[49] K. Y. Guslienko, Magnetic vortex state stability, reversal and dynamics in restricted geometries, J. Nanosci. Nanotechnol. **8**, 2745 (2008).

[50] J. Hajduček, A. Andrieux, J. A. Arregi, M. Tichý, P. Cattaneo, B. Ferrari, F. Carbone, V. Uhlíř, and T. LaGrange, Data for "Dislocation-driven nucleation type switching across repeated ultrafast magnetostructural phase transition", Zenodo (2025), https://doi.org/10.5281/zenodo.15624077.

[51] L. Piazza, D. J. Masiel, T. LaGrange, B. W. Reed, B. Barwick, and F. Carbone, Design and implementation of a fs-resolved transmission electron microscope based on thermionic gun technology, Chem. Phys. **423**, 79 (2013).

[52] HREM Research Inc., QPt module for DigitalMicrograph (HREM Research Inc., Tokyo, Japan, 2024), https://www.hremresearch.com/qpt/.

[53] J. D. Eshelby, The determination of the elastic field of an ellipsoidal inclusion, and related problems, Proc. R. Soc. London A **241**, 376 (1957).

[54] J. M. Lommel and J. S. Kouvel, Effects of mechanical and thermal treatment on the structure and magnetic transitions in FeRh, J. Appl. Phys. **38**, 1263 (1967).

[55] S. A. Makhlouf, T. Nakamura, and M. Shiga, Structure and magnetic properties of $FeAl_{1-x}Rh_x$ alloys, J. Magn. Magn. Mater. **135**, 257 (1994).

[56] L. Muldawer and F. deBergevin, Antiferromagnetic-ferromagnetic transformation in FeRh, J. Chem. Phys. **35**, 1904 (1961).

[57] L. Zsoldos, Lattice parameter change of FeRh alloys due to antiferromagnetic-ferromagnetic transformation, Phys. Stat. Sol. **20**, K25 (1967).

[58] A. Glavic and M. Björck, GenX 3: The latest generation of an established tool, J. Appl. Cryst. **55**, 1063 (2022).

[59] M. Richardson, D. Melville, and J. Ricodeau, Specific heat measurements on an FeRh alloy, Phys. Lett. A **46**, 153 (1973).

[60] D. Ourdani, A. Castellano, A. K. Vythelingum, J. A. Arregi, V. Uhlíř, B. Perrin, M. Belmeguenai, Y. Roussigné, C. Gourdon, M. J. Verstraete, and L. Thevenard, Experimental determination of the temperature- and phase-dependent elastic constants of FeRh, Phys. Rev. B **110**, 014427 (2024).

[61] M. Mattern, J. Pudell, J. A. Arregi, J. Zlámal, R. Kalousek, V. Uhlíř, M. Rössle, and M. Bargheer, Accelerating the laser-induced phase transition in nanostructured FeRh via plasmonic absorption, Adv. Funct. Mater. **34**, 2313014 (2024).